\documentclass[a4paper,11pt]{article}

\pdfoutput=1

\usepackage{jcappub_ver}
\usepackage{amsmath}
\usepackage{graphicx}

\usepackage[english]{babel}
\usepackage[utf8]{inputenc}
\usepackage{pdflscape}
\usepackage{enumerate}
\usepackage{amsbsy}
\usepackage{amsmath} 
\usepackage{graphics}
\usepackage{mathrsfs}
\usepackage{wrapfig}
\usepackage{mathtools}

\usepackage{amsfonts}
\usepackage{pstricks}
\usepackage{color}
\usepackage{setspace}
\usepackage{tensor}
\usepackage{braket}

\newcommand{\bea}{\begin{eqnarray}} \newcommand{\eea}{\end{eqnarray}}
\newcommand{\el}{\nonumber \\}
\newcommand{\re}[1]{(\ref{#1})}
\newcommand{\pat}{\partial}

\renewcommand{\sec}[1]{section \ref{#1}}
\newcommand{\fig}[1]{figure \ref{#1}}

\newcommand{\para}{\paragraph}

\renewcommand{\a}{\alpha}
\renewcommand{\b}{\beta}
\renewcommand{\c}{\gamma}
\renewcommand{\d}{\delta}

\newcommand{\ha}{\frac{1}{2}}

\newcommand{\rmd}{\mathrm{d}}

\newcommand{\ie}{i.e.\ }

\newcommand{\Mpl}{M_{{}_{\mathrm{Pl}}}}

\newcommand{\bo}{\bar\Omega}
\newcommand{\bh}{\bar h}
\newcommand{\ecm}{E}

\title{Tree-level unitarity in Higgs inflation in the metric and the Palatini formulation}

\author[a,b,c]{Asuka Ito,}
\author[d]{Wafaa Khater}
\author[e]{and Syksy R\"{a}s\"{a}nen}

\affiliation[a]{Department of Physics, Tokyo Institute of Technology, Tokyo 152-8551, Japan}
\affiliation[b]{Department of Physics, National Tsing-Hua University, Hsinchu 30013, Taiwan}
\affiliation[c]{Center for Theory and Computation, National Tsing-Hua University, Hsinchu 30013, Taiwan}

\affiliation[d]{Birzeit University, Department of Physics \\
P.O. Box 14, Birzeit, West Bank, Palestine}

\affiliation[e]{University of Helsinki, Department of Physics and Helsinki Institute of Physics,\\ P.O. Box 64, FIN-00014 University of Helsinki, Finland}

\emailAdd{asuka.i.aa@m.titech.ac.jp}
\emailAdd{wkhater@birzeit.edu}
\emailAdd{syksy.rasanen@iki.fi}

\abstract{
We calculate the tree-level amplitudes for electrically neutral $2\to2$ scattering for the Standard Model Higgs doublet non-minimally coupled to the Ricci scalar. We consider both the metric and the Palatini formulation of gravity. We find the partial wave unitarity limit for a general background field value. Our results are in agreement with previous work. In the electroweak vacuum, tree-level unitarity is violated at $\sim\Mpl/\xi$ in the metric formulation, and at $\sim\Mpl/\sqrt{\xi}$ in the Palatini formulation. In the inflationary large field background, the unitarity limit is at $\sim\Mpl/\sqrt{\xi}$ in both formulations. We compare the unitarity violation energy to scales relevant during inflation. We also calculate a direct collider limit on $\xi$ in the Palatini formulation, $\xi<2.5\times10^{31}$.
}

\begin{document}

\begin{flushleft}
	\hfill		 HIP-2021-32/TH
\end{flushleft}

\maketitle
  
\setcounter{tocdepth}{2}

\setcounter{secnumdepth}{3}

\section{Introduction} \label{sec:intro}

\para{Higgs inflation, unitarity, and the formulation of general relativity.}

Inflation is the most successful scenario for the primordial universe \cite{Starobinsky:1979ty, Starobinsky:1980te, Kazanas:1980tx, Guth:1981, Sato:1981, Mukhanov:1981xt, Linde:1981mu, Albrecht:1982wi, Hawking:1981fz, Chibisov:1982nx, Hawking:1982cz, Guth:1982ec, Starobinsky:1982ee, Sasaki:1986hm, Mukhanov:1988jd}. Perhaps the simplest inflationary model is the one where the Standard Model Higgs non-minimally coupled to the Ricci scalar $R$ is the inflaton \cite{Bezrukov:2007} (for reviews, see \cite{Bezrukov:2013, Bezrukov:2015, Rubio:2018}; for an earlier similar model, see \cite{Futamase:1987, Salopek:1988}). It involves only the degrees of freedom present in general relativity and the Standard Model, and the tree-level predictions are in excellent agreement with observations \cite{Akrami:2018odb}. Loop corrections can make a large difference if tuned, but for generic parameter values they have little impact on the predictions \cite{Espinosa:2007qp, Barvinsky:2008ia, Barvinsky:2009fy, Barbon:2009, Burgess:2009, Popa:2010xc, DeSimone:2008ei, Bezrukov:2008ej, Bezrukov:2009db, Barvinsky:2009ii, Bezrukov:2010, Bezrukov:2012sa, Allison:2013uaa, Salvio:2013rja, Shaposhnikov:2013ira, Weenink:2010, Calmet:2012eq, Steinwachs:2013tr, Bezrukov:2014bra, Hamada:2014iga, Prokopec:2014, Kamenshchik:2014waa, Burns:2016ric, Fumagalli:2016lls, Hamada:2016, Karamitsos:2017elm, Karamitsos:2018lur, George:2013iia, George:2015nza, Postma:2014vaa, Prokopec:2012, Herrero-Valea:2016jzz, Pandey:2016jmv, Pandey:2016unk, Bezrukov:2014ipa, Rubio:2015zia, Enckell:2016xse, Bezrukov:2017dyv, Rasanen:2017, Masina:2018ejw, Salvio:2017oyf, Ezquiaga:2017fvi, Rasanen:2018a, Enckell:2018a, Shaposhnikov:2020fdv, Fumagalli:2020ody, Enckell:2020lvn}. 

However, already at tree-level the non-minimal coupling $\xi$ of the Higgs doublet spoils the Standard Model cancellations between the Higgs and the $W$ and $Z$ bosons, leading to violation of tree-level perturbative unitarity at energies much below the Planck scale. The original Higgs inflation proposal used the metric formulation of general relativity, where tree-level unitarity violation in the electroweak (EW) vacuum occurs at $\sim\Mpl/\xi$ \cite{Barbon:2009, Burgess:2009, Bezrukov:2009db, Burgess:2010zq, Lerner:2009na, Lerner:2010mq, Hertzberg:2010, Bezrukov:2010, Atkins:2010yg, Bezrukov:2011a, Calmet:2013, Weenink:2010, Lerner:2011it, Xianyu:2013, Prokopec:2014, Ren:2014, Escriva:2016cwl, Fumagalli:2017cdo, Mikura:2021clt, Bezrukov:2013, Rubio:2018}. The conclusion can be changed by a $R^2$ term in the action, expected to be generated by renormalisation \cite{Barbon:2015, Salvio:2015kka, Salvio:2017oyf, Kaneda:2015jma, Calmet:2016fsr, Wang:2017fuy, Ema:2017rqn, Pi:2017gih, He:2018gyf, Gorbunov:2018llf, Ghilencea:2018rqg, Wang:2018kly, Gundhi:2018wyz, Karam:2018, Kubo:2018, Enckell:2018c, Ema:2019, Canko:2019mud, Enckell:2018a, He:2020ivk, Bezrukov:2020txg, Hill:2021psc, Antoniadis:2021axu, Ferreira:2021ctx, Mikura:2021clt}, direct coupling between the curvature and the Higgs kinetic term \cite{Germani:2010gm, Germani:2010ux, Kamada:2010qe, Atkins:2010yg, Kamada:2012se, Kamada:2013bia, Germani:2014hqa, Escriva:2016cwl, Fumagalli:2017cdo, Sato:2017qau, Granda:2019wip, Sato:2020ghj, Fumagalli:2020ody, Gialamas:2020vto, Karydas:2021wmx}, or modifications that involve higher derivatives of the Higgs field \cite{Koshelev:2020fok}. It is possible that unitarity is perturbatively restored beyond tree-level while keeping to the vanilla action, where the only ingredient added to the Standard model plus the Einstein--Hilbert action is the non-minimal coupling of the Higgs field \cite{Aydemir:2012nz, Calmet:2013}. Also, scatterings depend on the background field value, which may lead to a higher unitarity violation scale in the inflationary background \cite{Bezrukov:2009db, Bezrukov:2010, Bezrukov:2011a, Bezrukov:2013, Rubio:2018}.

Furthermore, even with the vanilla action and at tree-level, the results depend on the choice of the formulation of general relativity. There are several formulations, such as the metric formulation, the Palatini formulation \cite{Einstein:1925, Hehl:1976, Hehl:1978, Papapetrou:1978, Hehl:1981, ferraris1982, Percacci:1991, Percacci:2009, Percacci:2020bzf} and the teleparallel formulation \cite{Einstein:1928a, Einstein:1928b, Einstein:1930, Unzicker:2005, Aldrovandi:2013, Golovnev:2018red, Krssak:2018ywd}, which agree for the Einstein--Hilbert action with minimally coupled matter. However, the non-minimal coupling of the Higgs breaks the equivalence between different formulations. In particular, the inflationary predictions are different in the metric formulation, the Palatini formulation \cite{Bauer:2008, Bauer:2010, Rasanen:2017, Enckell:2018a, Markkanen:2017, Rasanen:2018a, Rasanen:2018b, Rubio:2019, Raatikainen:2019, Langvik:2020, McDonald:2020, Shaposhnikov:2020fdv, Tenkanen:2020dge, Langvik:2020, Enckell:2020lvn, Shaposhnikov:2020frq, Antoniadis:2021axu, Mikura:2021clt}, and the teleparallel formulation \cite{Raatikainen:2019}. In the Palatini case, unitarity is expected to be violated at energies $\sim\Mpl/\sqrt{\xi}$, parametrically higher than in the metric formulation, although in the Palatini case $\xi$ is typically larger \cite{Bauer:2010, McDonald:2020, Shaposhnikov:2020fdv, Enckell:2020lvn, Antoniadis:2021axu, Mikura:2021clt}.

The tree-level $2\to2$ amplitudes for electrically neutral scattering for the Higgs doublet and the resulting unitarity bound have been calculated in the metric formulation in \cite{Xianyu:2013, Ren:2014}, and one scattering amplitude was calculated in both the metric and the Palatini formulation in \cite{Antoniadis:2021axu}. We extend the calculations, treating the metric formulation and the Palatini formulation on an equal footing, and considering all neutral $2\to2$ Higgs doublet scatterings.

In \sec{sec:calc} we write down the action and expand up to the four-field couplings for the excitations of the Higgs doublet fields in a general constant background field value. In section \ref{sec:uni} we calculate the partial wave amplitudes for $2\to2$ scattering and apply the partial wave unitarity condition. We also extend a calculation of a direct upper bound on $\xi$ from colliders to the Palatini case. Our conclusions about the unitarity violation scale in both the metric and the Palatini formulation agree with previous work. In the metric case, tree-level unitarity is violated at $\sim\Mpl/\xi$ in the EW vacuum, and at $\sim\Mpl/\sqrt{\xi}$ in the inflationary large field background. In the Palatini case, the tree-level unitarity violation scale is $\sim\Mpl/\sqrt{\xi}$ in both cases. We compare the unitarity violation scale to the Hubble rate, the energy density, and the background field values, contrasting the metric and the Palatini formulation. We summarise our findings and outline open issues in \sec{sec:conc}.

\section{Action} \label{sec:calc}

\subsection{Jordan frame}

We consider the Standard Model plus the Einstein--Hilbert action supplemented with a coupling between the Higgs doublet $H$ and the Ricci scalar,
\bea \label{actionJ}
  S &=& \int \rmd^4 x \sqrt{-g} \bigg[ \left( \frac{M^2}{2} + \xi H^\dagger H \right) g^{\a\b} R_{\a\b} - g^{\a\b} \left( D_\a H \right)^{\dagger}  D_\b H - V(H) \el
  && - \frac{1}{4} \sum_{a} g^{\a\c} g^{\b\d} W^{(a)}_{\a\b} W^{(a)}_{\c\d} - \frac{1}{4} g^{\a\c} g^{\b\d} B_{\a\b} B_{\c\d} + \sum_{n} i \bar{\Psi}^{(n)} \gamma^A e_A{}^\a \mathcal{D}_\a \Psi^{(n)} \bigg] \ ,
\eea
where $g$ is the determinant of the metric $g_{\a\b}$, $R_{\a\b}$ is the Ricci tensor, $M$ is a mass scale and $\xi$ is a positive constant. In the metric formulation, $R_{\a\b}$ is determined by the metric via the Levi--Civita connection, in the Palatini formulation it depends only on the connection and is independent of the metric. We take $M=\Mpl$ (see \cite{Rasanen:2018b} for discussion of this choice), and use Planck units, $\Mpl=1$. The tensors $W^{(a)}_{\a\b}$ and $B_{\a\b}$ are the field strength of the $SU(2)_{L}$ gauge field $W^{(a)}_{\a}$ and of the $U(1)_{Y}$ gauge field $B_{\a}$, respectively. The Higgs gauge covariant derivative is $D_{\a} = \partial_\a + i g_{2} \sum_{a}W^{(a)}_{\a}\frac{\sigma_{a}}{2} + i\frac{g_{1}}{2}B_{\a}$, where $g_1$ and $g_2$ are the gauge couplings. The term $\Psi^{(n)}$ contains the quarks and leptons, $\gamma^A$ are the gamma matrices, $e_A{}^\a$ is the inverse tetrad, $\mathcal{D}_{\a} = \partial_{\a} - \frac{i}{2} \omega_\a{}^{AB} \sigma_{AB} + i g_{2} \sum_{a} W^{(a)}_{\a}\frac{\sigma_{a}}{2} + i\frac{g_{1}}{2}B_{\a}$, where $\omega_\a{}^{AB}= - e^A{}_\b g^{\b\c} \partial_\a e^B{}_\c + e^A{}_\c e^B{}_\b g^{\b\d} \Gamma^\c_{\d\a}$ is the spin connection, $\sigma_{AB} = \frac{i}{4} [ \gamma_A, \gamma_B ]$, and $\Gamma^\c_{\a\b}$ is the affine connection. We have omitted the Yukawa interactions, as their contribution is subdominant to the derivative interactions we consider. Even without the Yukawa term, in the Einstein frame the fermions will couple to the Higgs at tree-level. The Higgs potential is
\bea \label{V}
  V(H) &=& \lambda \left(H^\dagger H - \ha v^2\right)^2 \ ,
\eea
where $v=246$ GeV is the EW vacuum expectation value.

In the Palatini formulation the fermions (in addition to the non-minimal Higgs coupling) contribute a source term to the connection equation of motion, unless they are coupled to the Levi--Civita spin connection instead of the full spin connection, or have a non-minimal kinetic term \cite{Randono:2005, Freidel:2005sn, Mercuri:2006wb, Bojowald:2007nu, Shaposhnikov:2020gts, Shaposhnikov:2020aen}. We neglect this effect, which is subdominant to the couplings we consider.

\subsection{Einstein frame}

It is easier to calculate when the Higgs is minimally coupled, so we shift to the Einstein frame with the Weyl transformation
\bea \label{weyl}
  g_{\a\b} \to \Omega^{-2} g_{\a\b} \ ,
\eea
where we choose
\bea \label{omega}
   \Omega^{2} = 1 + 2 \xi H^{\dagger}H \ .
\eea
The tetrad and the connection transform as
\bea
  e^A{}_\a &\to& \Omega^{-1} e^A{}_\a \ ,  \label{f1}  \\
  \Gamma^{\c}_{\a\b} &\to& \Gamma^{\c}_{\a\b}
       - p ( \delta^{\c}{}_{\a} \pat_\b \ln\Omega + \delta^{\c}{}_{\b} \pat_\a \ln\Omega - g_{\a\b} g^{\c\d} \pat_\d \ln\Omega ) \ ,  \label{f2}
\eea
where $p=1$ in the metric formulation and $p=0$ in the Palatini formulation. Hence the spin connection transforms as
\bea
  \omega_\a{}^{AB} &\to& \omega_\a{}^{AB} + ( 1 - p ) \eta^{AB} \pat_\a \ln\Omega - 2 p e^{[A}{}_\a e^{B]}{}_\b g^{\b\c} \pat_\c \ln\Omega \ .  \label{f3}
\eea
 The Ricci scalar term in the action changes as
\begin{equation} \label{ricci}
   \Omega^2 \sqrt{-g} g^{\a\b} R_{\a\b} \to \sqrt{-g} g^{\a\b} \left( 
                R_{\a\b} + 3 p \nabla_\a \nabla_\b \ln\Omega^{2} 
                - p \frac{3}{2} \pat_\a \ln\Omega^{2}  
                                                 \pat_\b \ln\Omega^{2} 
              \right) \ , 
\end{equation}
where the second term is a total derivative in the metric case (and zero in the Palatini case), so we drop it. With (\ref{weyl})--(\ref{ricci}), the action \re{actionJ} becomes
\bea \label{actionE}
  S &=& \int \rmd^4 x \sqrt{-g} \bigg\{ \ha g^{\a\b} R_{\a\b} - \frac{1}{\Omega^2} g^{\a\b} \left( D_\a H \right)^{\dagger}  D_\b H - p \frac{3 \xi^{2}}{\Omega^4} g^{\a\b} \pat_\a (H^{\dagger}H) \pat_\b (H^{\dagger}H) \el
  && - U(H) - \frac{1}{4} \sum_{a} g^{\a\c} g^{\b\d} W^{(a)}_{\a\b} W^{(a)}_{\c\d} - \frac{1}{4} g^{\a\c} g^{\b\d} B_{\a\b} B_{\c\d} \el
  && + \sum_{n} i \bar{\Psi}^{(n)} \gamma^A \left[ \frac{1}{\Omega^3} e_A{}^\a \mathcal{D}_\a + p \frac{\xi}{\Omega^5} i \sigma_{AB} e^B{}_\b g^{\b\c} \pat_\c (H^{\dagger} H) \right] \Psi^{(n)} \bigg\} \ ,
\eea
where we have denoted $U\equiv V/\Omega^4$. The transformation to the Einstein frame has shifted the effect of the Higgs non-minimal coupling from the Higgs-gravity sector to the Higgs and fermion sectors. We drop the gauge fields, as their kinetic terms are invariant in the Weyl transformation, and hence do not pick up any factors of $\xi$, and their coupling with the Higgs is not enhanced by positive powers of $\xi$. Similarly, the potential only contributes subdominant terms that are not enhanced by $\xi$ and do not grow with energy, so we neglect its effect on scattering. Analysis of unitarity violation in terms of the potential and the background field (which has only one physical component) can be misleading, as the leading order unitarity violation is related to the curvature of field space \cite{Burgess:2010zq}, which is absent in the single field case \cite{Lerner:2009na, Burgess:2010zq, Hertzberg:2010, Lerner:2011it}. (On Higgs inflation and unitarity from the point of view of field space curvature, see \cite{Karananas:2016kyt, Rubio:2018, Karananas:2020qkp, Mikura:2021clt}.)

\subsection{Action for excitations} \label{sec:excitation}

Let us now derive tree-level Higgs $2\to2$ scattering amplitudes for the action \re{actionE} and consider partial wave unitarity \cite{Itzykson:1980rh}. We parametrise the Higgs doublet as
\begin{equation} \label{dob2}
H =
\begin{pmatrix}
  \pi^{+} \\
  \frac{1}{\sqrt{2}} ( \bh + h + i \pi^{0} )
\end{pmatrix} \ ,
\end{equation}
where $\bh$ is the time-dependent vacuum expectation value of the Higgs field, $h$ is the physical Higgs field, and the complex scalar field $\pi^{+}=(\pi^{-})^{\dagger}$ and the real scalar field $\pi^{0}$ are Goldstone bosons. We have $H^\dagger H=\ha\bh^2 + \bh h + \ha h^2 + \ha (\pi^0)^2 + \pi^{+} \pi^{-}$, so the conformal factor \re{omega} reads
\begin{equation} \label{omegaexp}
  \Omega^{2} = \bo^2 
           \left[  1 + \frac{\xi}{\bo^2}
                 \left( 2 \bh h + h^2 + \pi^2 \right)   \right] \ ,
\end{equation}
where $\bo^2\equiv1+\xi{\bar h}^2$ is the background value, and we have denoted $\pi^2\equiv(\pi^0)^2+2\pi^{+}\pi^{-}$. We insert \re{dob2} and \re{omegaexp} into the action \re{actionE} and expand to quartic order in the field excitations, which includes all vertices appearing in tree-level $2\to2$ scattering. In the Einstein frame, we can neglect gravitons, unlike in the Jordan frame \cite{Ren:2014}. We do not make any expansion in the background field. The resulting action is
\bea \label{actionE2}
  S &=& \int \rmd^4 x \bigg\{ - \ha \frac{ \bo^2 + p 6 \xi^2 \bh^2 }{\bo^4} \eta^{\a\b} \pat_\a h \pat_\b h - \ha \frac{1}{\bo^2} \eta^{\a\b} \pat_\a \pi^0 \pat_\b \pi^0 - \frac{1}{\bo^2} \eta^{\a\b} \pat_\a \pi^+ \pat_\b \pi^- \el
 && + \frac{\xi}{\bo^4} \left( \bh h + \frac{1 - 3 \xi \bh^2}{2 \bo^2} h^2 + \ha \pi^2 \right) \eta^{\a\b} ( \pat_\a h \pat_\b h + \pat_\a \pi^0 \pat_\b \pi^0 + 2 \pat_\a \pi^+ \pat_\b \pi^- ) \el
 && + p \frac{3 \xi^2}{4 \bo^4} ( 4 \bh h + h^2 + \pi^2 ) \eta^{\a\b} \pat_\a \pat_\b ( h^2 + \pi^2 ) - p \frac{6 \xi^3 \bh^2}{\bo^6} \pi^2 \eta^{\a\b} \pat_\a \pat_\b h^2 \el
  && + p \frac{6 \xi^3 \bh^2}{\bo^6} \left( 2 \bh h + \frac{5 - \xi \bh^2}{\bo^2} h^2 + \pi^2 \right) \eta^{\a\b} \pat_\a h \pat_\b h \el
  && + \sum_{n} i \bar{\Psi}^{(n)} \frac{1}{\bo^3} \gamma^A \bigg[ \left( 1 - \frac{3 \xi \bh}{\bo^2} h + \xi \frac{- 3 + 12 \xi \bh^2}{2\bo^4} h^2 - \frac{3 \xi}{2\bo^2} \pi^2 \right) e_A{}^\a \mathcal{D}_\a \el
  && + p \frac{\xi}{\bo^2} i \sigma_{AB} e^B{}_\b \eta^{\b\c} \pat_\c \left( \bh h + \frac{1 -  4\xi \bh^2}{2 \bo^2} h^2 + \ha \pi^2 \right) \bigg] \Psi^{(n)} \bigg\} \ .
\eea
We have assumed that the spacetime can be approximated as Minkowski space, \ie we consider scattering energies much higher than the Hubble rate. We also neglect time and space derivatives of $\bh$. This approximation may not hold during preheating, when $\bh$ can vary rapidly in time and space.

We change variables to obtain canonical kinetic terms:
\bea \label{can}
  h \rightarrow \frac{\bo^2}{\sqrt{\bo^2 + p 6 \xi^2 \bh^2}} h \ , \quad \pi^{i} \rightarrow \bo \pi^{i} \ , \quad \Psi^{(n)} \rightarrow \bo^{3/2} \Psi^{(n)} \ .
\eea
With this transformation, the action \re{actionE2} becomes
\bea \label{actionEcan}
  S &=& \int \rmd^4 x \bigg\{ - \ha \eta^{\a\b} \pat_\a h \pat_\b h - \ha \eta^{\a\b} \pat_\a \pi^0 \pat_\b \pi^0 - \eta^{\a\b} \pat_\a \pi^+ \pat_\b \pi^- \el
    && + \xi \left[ \frac{\bh}{\sqrt{\bo^2 + p 6 \xi^2 \bh^2}} h + \frac{1 - 3 \xi \bh^2}{2 ( \bo^2 + p 6 \xi^2 \bh^2 )} h^2 + \ha \pi^2 \right] \times \el
    && \times \eta^{\a\b} \left( \frac{\bo^2}{\bo^2 + p 6 \xi^2 \bh^2} \pat_\a h \pat_\b h + \pat_\a \pi^0 \pat_\b \pi^0 + 2 \pat_\a \pi^+ \pat_\b \pi^- \right) \el
    && + p \frac{3}{4} \xi^2 \left( \frac{4 \bh}{\sqrt{\bo^2 + 6 \xi^2 \bh^2}} h + \frac{\bo^2}{\bo^2 + 6 \xi^2 \bh^2} h^2 + \pi^2 \right) \eta^{\a\b} \pat_\a \pat_\b \left( \frac{\bo^2}{\bo^2 + 6 \xi^2 \bh^2} h^2 + \pi^2 \right) \el
  && - p \frac{6 \xi^3 \bh^2}{\bo^2 + 6 \xi^2 \bh^2} \pi^2 \eta^{\a\b} \pat_\a \pat_\b h^2 \el
  && + p \frac{6 \xi^3 \bh^2}{\bo^2 + 6 \xi^2 \bh^2} \left( \frac{2 \bh}{\sqrt{\bo^2 + 6 \xi^2 \bh^2}} h + \frac{ 5 - \xi \bh^2 }{\bo^2 + 6 \xi^2 \bh^2} h^2 + \pi^2 \right) \eta^{\a\b} \pat_\a h \pat_\b h \el
   && + \sum_{n} i \bar{\Psi}^{(n)} \gamma^A \bigg[ \left( 1 - \frac{3 \xi \bh}{\sqrt{\bo^2 + p 6 \xi^2 \bh^2}} h + \xi \frac{- 3 + 12 \xi \bh^2}{2 ( \bo^2 + p 6 \xi^2 \bh^2 )} h^2 - \frac{3}{2} \xi \pi^2 \right) e_A{}^\a \mathcal{D}_\a \el
  && + p \xi \sigma_{AB} e^B{}_\b \eta^{\b\c} \pat_\c \left( \frac{\bh}{\sqrt{\bo^2 + 6 \xi^2 \bh^2}} h + \frac{1 -  4\xi \bh^2}{2 (\bo^2 + 6 \xi^2 \bh^2)} h^2 + \ha \pi^2 \right) \bigg] \Psi^{(n)} \bigg\} \ .
\eea
In the Palatini case, this action is rather simple:
\bea \label{actionEcanPal}
  S &=& \int \rmd^4 x \bigg\{ - \ha \eta^{\a\b} \pat_\a h \pat_\b h - \ha \eta^{\a\b} \pat_\a \pi^0 \pat_\b \pi^0 - \eta^{\a\b} \pat_\a \pi^+ \pat_\b \pi^- \el
    && + \xi \left( \frac{\bh}{\bo} h + \frac{1 - 3 \xi \bh^2}{2 \bo^2} h^2 + \ha \pi^2 \right) \eta^{\a\b} \left( \pat_\a h \pat_\b h + \pat_\a \pi^0 \pat_\b \pi^0 + 2 \pat_\a \pi^+ \pat_\b \pi^- \right) \el
  && + \sum_{n} i \bar{\Psi}^{(n)} \gamma^A \left( 1 - \frac{3 \xi \bh}{\bo} h + \xi \frac{- 3 + 12 \xi \bh^2}{2 \bo^2} h^2 - \frac{3}{2} \xi \pi^2 \right) e_A{}^\a \mathcal{D}_\a \Psi^{(n)} \bigg\} \ .
\eea
In the metric case, the Higgs-Goldstone sector contains derivative terms that grow like $\xi^2$ for $\xi\gg1$, the relevant limit for inflation, while the Higgs-fermion terms only grow like $\xi$. The terms that involve fermions can therefore be neglected when calculating $2\to2$ tree-level scattering. In contrast, in the Palatini case both the leading Higgs-Goldstone terms and the Higgs-fermion terms are proportional to $\xi$, so there is no obvious justification for neglecting the fermion terms. We consider only the Higgs-Goldstone interactions, leaving the interactions involving fermions for future work. When calculating the $2\to2$ scattering amplitudes, we utilised the longitudinal-Goldstone boson equivalence theorem \cite{Chanowitz:1985hj}, where the high-energy amplitudes of longitudinal physical vector gauge bosons are well approximated by amplitudes of the corresponding unphysical Goldstone scalar bosons.

The new Higgs coupling terms with extra derivatives spoil the Standard Model cancellation in scattering amplitudes between the Higgs and the weak gauge bosons, leading to amplitudes that grow like $E^2$ and are proportional to either $\xi^2$ or $\xi$ (depending on the formulation), implying violation of tree-level unitarity for energies $E$ much smaller than Planck scale. Let us look at this in detail.

\section{Unitarity bound} \label{sec:uni}

\subsection{Partial wave amplitudes}

We consider two particles with momenta $p_1, p_2$ scattering into two particles with momenta $p_1', p_2'$ in the centre of mass frame. The $S$-matrix element is
\begin{equation}
  \braket{p_{1}',p_{2}'|p_{1},p_{2}} = (2\pi)^{4} \delta^{(4)}(p_{1} + p_{2} - p_{1}' -p_{2}') i T  \ .
  \label{delta}
\end{equation}
We obtain the following scattering amplitudes from the action \re{actionEcan}:
\begin{eqnarray} \label{T}
  T[\pi^{+}\pi^{-} \rightarrow \pi^{+}\pi^{-}] &=& \ha [ 2 ( 1 + p 3\xi ) - ( 1 + p 6 \xi )^2 A ] \xi ( 1 - \cos\theta ) \ecm^2 \el
  T[\pi^{+}\pi^{-} \rightarrow \pi^{0}\pi^{0}] &=& \frac{1}{\sqrt{2}} \left[ 2 ( 1 + p 3\xi ) - ( 1 + p 6 \xi )^2 A \right] \xi \ecm^{2} \el
  T[\pi^{+}\pi^{-} \rightarrow h h] &=& \frac{1}{\sqrt{2}} \left[ 2 C + ( 1+ p 6 \xi ) B + A \right] \xi \ecm^2 \el
  T[\pi^{0}\pi^{0} \rightarrow h h] &=& \frac{1}{2}
      \left[ 2 C + ( 1 + p 6 \xi ) B +  A \right] \xi \ecm^{2} \el
  T[\pi^{0} h \rightarrow \pi^{0} h] &=& -\frac{1}{2}
     \left[ 2 C + ( 1 + p 6 \xi ) B + A \right] \xi ( 1 - \cos\theta ) \ecm^2 \ ,
\end{eqnarray}
where $\ecm$ is the center of mass energy, $\theta$ is the angle between $\vec{p}_{1}$ and $\vec{p}_{1}'$, and
\bea
  A &\equiv& \frac{x^2}{1 + p 6 \xi x^2} \el
  B &\equiv& \frac{x^2}{(1 + p 6 \xi x^2)^2} [ 1 + p \xi ( - 6  + 12 x^2 ) ] \el
  C &\equiv& \frac{ 1+ p 3 \xi}{1 + p 6 \xi x^2} ( 1 - 2 x^2 ) \ ,
\eea
with $x\equiv\sqrt{\xi}\bh/\bo$. We have neglected particle masses as small compared to $E$. We have included factors of $1/\sqrt{2}$ for initial and final states with identical particles. The matrix elements for $hh\to hh$ and $\pi^0\pi^0\to\pi^0\pi^0$ scattering vanish due to crossing symmetry (they receive small $\mathcal{O}(E^0)$ contributions from the potential).

In \cite{Xianyu:2013} the amplitudes were calculated in the EW vacuum in the metric formulation to leading order in $\xi v^2$. In \cite{Ren:2014}, scattering amplitudes were calculated in a general field background in the metric formulation. Our results fully agree with those of \cite{Xianyu:2013, Ren:2014}. In \cite{Antoniadis:2021axu} the $\pi^0 h\to\pi^0 h$ scattering amplitude was calculated in the large field limit in both the metric and the Palatini case. Our results disagree with those of \cite{Antoniadis:2021axu} both as regards numerical factors and the dependence on $\xi$.
 
To obtain the unitarity constraint, we expand the matrix elements in terms of partial waves,
\begin{equation}
  T(\ecm, \theta) = 16\pi \sum_{l=0}^{\infty} (2l+1) P_{l}(\cos\theta) a_{l}(\ecm) \ ,
\end{equation}
where
\begin{equation} \label{par}
  a_{l}(\ecm) = \frac{1}{32\pi} \int^{1}_{-1} \rmd(\cos\theta) P_{l}(\cos\theta) T(\ecm, \theta)
\end{equation}
is the partial wave scattering matrix element, and $P_l$ are Legendre polynomials. We then apply the partial wave unitarity condition \cite{DiLuzio:2016sur}
\begin{equation}
  |{\rm Re}(a_{0})| < \frac{1}{2} \ .  \label{puc}
\end{equation}
From \re{T} and \re{par}, we obtain the $l=0$ partial wave modes
\begin{eqnarray} \label{a0}
  a_0[\pi^{+}\pi^{-} \rightarrow \pi^{+}\pi^{-}] &=& \frac{1}{32\pi} [ 2 (1 + p 3\xi ) - ( 1 + p 6 \xi )^2 A ] \xi \ecm^2 \el
  a_0[\pi^{+}\pi^{-} \rightarrow \pi^{0}\pi^{0}] &=& \frac{1}{16\pi \sqrt{2}} \left[ 2 ( 1 + p 3\xi ) - ( 1 + p 6 \xi )^2 A \right] \xi \ecm^{2} \el
  a_0[\pi^{+}\pi^{-} \rightarrow h h] &=& \frac{1}{16\pi \sqrt{2}} \left[ 2 C + ( 1+ p 6 \xi ) B + A \right] \xi \ecm^2 \el
  a_0[\pi^{0}\pi^{0} \rightarrow h h] &=& \frac{1}{32\pi} \left[ 2 C +  ( 1 + p 6 \xi ) B + A \right] \xi \ecm^{2} \el
  a_0[\pi^{0} h \rightarrow \pi^{0} h] &=&-\frac{1}{32\pi} \left[ 2 C + ( 1 + p 6 \xi ) B + A  \right] \xi \ecm^2 \ .
\end{eqnarray}

\subsection{Electroweak vacuum} \label{sec:EW}

In the EW vacuum $\bar{h} = v = 246$ GeV, so $\xi \bh^2\ll1$, and $A,B\to0$, $C\to1+p3\xi$. In the basis $\{\ket{\pi^{+}\pi^{-}}, \frac{1}{\sqrt{2}} \ket{\pi^{0}\pi^{0}}, \frac{1}{\sqrt{2}}\ket{hh}, \ket{\pi^{0}h}\}$ the scattering amplitudes \re{a0} reduce to the following matrix:
\begin{equation}
  a_{0}(\ecm) = \frac{( 1 + p 3\xi ) \xi}{16\pi} \ecm^{2}
      \begin{pmatrix}
            1 & \sqrt{2} & \sqrt{2} & 0 \\
            \sqrt{2} & 0 & 1 & 0 \\
            \sqrt{2} & 1 & 0 & 0 \\
            0 & 0 & 0 & -1
      \end{pmatrix}  \ .  \label{pwcon}
\end{equation}
In the metric case, this leading order result was given in \cite{Xianyu:2013}. The metric formulation result in a general background was given in \cite{Ren:2014}, and expanded around the EW vacuum to the next-to-leading order $\mathcal{O}(\xi^4 v^2)$. As noted above, our results agree with \cite{Xianyu:2013, Ren:2014}.

Diagonalising the numerical part of the matrix gives ${\rm diag}(3,-1,-1,-1)$, so the unitarity bound (\ref{puc}) reads
\begin{equation}
  \ecm^{2} < \frac{8\pi}{ ( 1 + p 3\xi ) 3 \xi} \ . \label{ineq}
\end{equation}
We denote the scattering energy that saturates this limit by $\Lambda$. In the metric formulation ($p=1$), for $\xi\gg1$ we get $\Lambda=\sqrt{\frac{8\pi}{9}}\xi^{-1}\sim\xi^{-1}$ \cite{Barbon:2009, Xianyu:2013, Ren:2014}. In the Palatini formulation ($p=0$), the tree-level unitarity violation scale is $\Lambda=\sqrt{\frac{8\pi}{3}}\xi^{-1/2}\sim\xi^{-1/2}$, as argued in \cite{Bauer:2010}.

If we naively apply the partial wave unitarity bound (\ref{pwcon}) to collider experiments by demanding that $\Lambda<1$ TeV, we get the upper bound $\xi<4\times10^{15}$ in the metric case and $\xi<0.7\times10^{32}$ in the Palatini case. This gives a pretty good approximation to the more careful analysis based on non-derivative interactions leading to Higgs production and decay done in \cite{Atkins:2012yn} for the metric case, and which we now update and extend to the Palatini case. In \cite{Atkins:2012yn} the $2\sigma$ constraint $\xi< 2.6\times10^{15}$ in the metric formulation was derived using data on the global Higgs branching ratio from the ATLAS and CMS experiments at LHC; updated limits were given in \cite{Xianyu:2013, Ren:2014, Wu:2019hso}.

The canonical background inflaton field $\bar\chi$ is related to $\bh$ by \cite{Bezrukov:2007, Bauer:2008}
\bea \label{dchi}
  \frac{\rmd\bar\chi}{\rmd\bh} = \sqrt{\frac{1+(1+p6\xi)\xi\bh^2}{(1+\xi \bh^2)^2}} \ ,
\eea
In \cite{Atkins:2012yn}, the Higgs field in Standard Model interaction vertices was written as $h(\chi)\equiv\kappa\chi$, where $\kappa=\frac{\rmd\bh}{\rmd\bar\chi}$. In the EW vacuum $\bh=v$, and we assume that $\xi\gg1$ and $\xi v^2\ll1$. If the first condition is not satisfied, the effects at collider scales are negligible; if the second condition is not satisfied, the effects are too large to agree with observation. We have
\begin{equation}
  \kappa = \frac{1+\xi v^2}{\sqrt{1+\xi v^2 + p 6 \xi^2 v^2}}
       \simeq 
\left\{
\begin{array}{ll}
  \left(  1 + 6\xi^{2}v^{2}  \right)^{-1/2} \quad &{\rm metric} \\
  \left(  1+\xi v^2  \right)^{1/2} \quad &{\rm Palatini}  \ ,
\end{array}
\right.
\end{equation}
where we have dropped a subleading term in the metric case. The field amplitude is decreased in the metric formulation and increased in the Palatini formulation. The cross section of Higgs production and decay is multiplied by a factor of $\kappa^{2}$ in the narrow width approximation. The current collider limit on the Higgs global signal strength is $1.13\pm0.06$ (at 1$\sigma$) \cite{Zyla:2020zbs}. This means that in the metric case, any value $\xi>0$ is ruled out at $2\sigma$; it is more informative to quote the $3\sigma$ limit $\xi<0.93\times10^{15}$. In the Palatini case, we have $\xi<2.5\times10^{31}$ at $2\sigma$ and $\xi<3.0\times10^{31}$ at $3\sigma$. The collider bound is much weaker in the Palatini formulation, because $\kappa\propto\xi$, instead of the metric formulation dependence $\kappa\propto\xi^2$.

The violation of tree-level unitarity in the EW vacuum implies one of three possibilities. Unitarity may be restored by 1) higher order perturbative effects in the theory studied \cite{Aydemir:2012nz, Calmet:2013}, 2) non-perturbative effects in the theory studied, or 3) new physics not included in the theory studied. In case 3) we would expect to see the new physics explicitly around and above the unitarity violation scale: $\Lambda\sim1/\xi$ in the metric case and $\Lambda\sim1/\sqrt{\xi}$ in the Palatini case. As tree-level inflation happens at field values $>1/\sqrt{\xi}$ in the metric case and $>1$ in the Palatini case, it has been argued that the new physics must therefore be included to describe inflation \cite{Barbon:2009, Burgess:2009, Burgess:2010zq, Hertzberg:2010}. However, in case 1) the perturbative scattering amplitudes may depend on the background field value in such a way that during inflation unitarity violation is only reached for scales higher than the inflationary scale, and likewise in case 2) the strong coupling scale of the theory may be different in different backgrounds \cite{Bezrukov:2009db, Bezrukov:2010, Bezrukov:2011a, Bezrukov:2013, Rubio:2018}. Let us now consider the amplitudes in the inflationary region.

\subsection{Inflationary region} \label{higinf}

During inflation $\xi \bh^2\gg1$, so $x\to1$, and $A,B\to1/(1 + p 6 \xi)$, $C\to-(1+p3\xi)/(1 + p 6 \xi)$. The partial wave amplitudes \re{a0} reduce to
\begin{eqnarray} \label{a0large}
  a_0[\pi^{+}\pi^{-} \rightarrow \pi^{+}\pi^{-}] &=& \frac{1}{32\pi} \xi \ecm^2 \el
  a_0[\pi^{+}\pi^{-} \rightarrow \pi^{0}\pi^{0}] &=& \frac{1}{16\pi \sqrt{2}} \xi \ecm^{2} \el
  a_0[\pi^{+}\pi^{-} \rightarrow h h] &=& - \frac{1}{8\pi\sqrt{2}} \frac{1}{(1 + p 6 \xi) \xi \bh^2} \xi \ecm^2 \el
  a_0[\pi^{0}\pi^{0} \rightarrow h h] &=& \frac{1}{16\pi} \frac{1}{(1 + p 6 \xi) \xi \bh^2} \xi \ecm^2 \el
  a_0[\pi^{0} h \rightarrow \pi^{0} h] &=& - \frac{1}{16\pi} \frac{1}{(1 + p 6 \xi) \xi \bh^2} \xi \ecm^2 \ .
\end{eqnarray}
As noted above, our metric formulation amplitudes agree with those in \cite{Ren:2014}, and our metric and Palatini results disagree with \cite{Antoniadis:2021axu}, where the $\pi^{0} h \rightarrow \pi^{0} h$ scattering amplitude was calculated. The difference with \cite{Antoniadis:2021axu}, although large for the amplitude of  the scattering $\pi^{0} h \rightarrow \pi^{0} h$, does not lead to a significant difference in $\Lambda$, because in our calculation the contribution of this channel is negligible, and the amplitudes of the dominant channels have the same parametric dependence on $\xi$ as the \cite{Antoniadis:2021axu} amplitude for $\pi^{0} h \rightarrow \pi^{0} h$ scattering.

In the limit $\xi\gg1$, the dominant scattering amplitudes in \re{a0large} grow like $\xi E^2$, in both the metric and the Palatini formulation. In order to get the precise unitarity violation scale $\Lambda$ for a general field value, we diagonalise the scattering matrix \re{a0} and apply the partial wave unitarity condition \re{puc} to the largest eigenvalue. We compare this scale to the Hubble rate $H$\footnote{Not to be confused with the Higgs doublet introduced in \sec{sec:calc}.}, the energy density scale $U^{1/4}$, the background field $\bh$, and the canonical background field $\bar\chi$.

The scattering amplitudes \re{a0} depend on the two variables $\bar h$ and $\xi$, and the former appears only in the combination $\xi \bh^2$. The value of $\xi$ is related to the Higgs quartic coupling $\lambda$ by the amplitude $A_{\text{s}}$ of the inflationary scalar perturbations,
\bea \label{As}
  A_{\text{s}} &=& \frac{N^2}{12 \pi^2} \frac{\lambda}{\xi (1+p6\xi)} \ ,
\eea
where $N$ is the number of e-folds from the CMB pivot scale $k=0.05$ Mpc$^{-1}$
until the end of inflation, and $\lambda$ is the Higgs quartic coupling. The observed value is $A_{\text{s}}=2\times10^{-9}$ \cite{Akrami:2018odb}. The number of e-folds in the metric case is $N=55-\Delta N$, where $\Delta N$ is the duration of reheating. In the metric formulation it is not clear whether $\Delta N=4$ or $\Delta N\ll1$ \cite{Bezrukov:2008ut, GarciaBellido:2008ab, Figueroa:2009, Figueroa:2015, Repond:2016, Ema:2016, DeCross:2016, Sfakianakis:2018lzf, Hamada:2020kuy}. We assume $\Delta N=4$. In the Palatini formulation reheating is almost instant, and the number of e-folds depends on $\xi$ \cite{Rubio:2019}. Gathering both cases, we have
\bea \label{N}
  N &=& 51 \qquad\qquad\qquad\qquad\qquad\quad\ \, \text{metric} \el
  N &=& 55 - \frac{1}{4} \ln\xi = 49 - \frac{1}{4} \ln\lambda \qquad \text{Palatini} \ ,
\eea
where we have used \re{As}. We have approximately $\xi=4\times10^4\sqrt{\lambda}$ in the metric case and $\xi=1\times10^{10}\lambda$ in the Palatini case. The perturbativity limit $\lambda<1$ gives $\xi<4.3\times10^4$ in the metric formulation and $\xi<1\times10^{10}$ in the Palatini formulation, a much tighter constraint than the collider limit discussed in \sec{sec:EW}. The number of e-folds is related to the field value during slow-roll as \cite{Bezrukov:2007, Rasanen:2017}
\bea \label{Nh}
  N = \frac{1+p6\xi}{8} \bh^2 \ .
\eea

Integrating \re{dchi}, we obtain the canonical background inflaton field $\bar\chi$ in terms of $\bh$ (setting $\chi(h=0)=0$) \cite{Rasanen:2017}
\bea \label{chi}
  \sqrt{\xi} \bar\chi &=& \sqrt{1+p6\xi} \,\mathrm{arsinh}(\sqrt{(1+p6\xi)\xi} \bh) - p \sqrt{6\xi} \, \mathrm{artanh} \frac{\sqrt{6} \xi \bh}{\sqrt{1+(1+6\xi)\xi \bh^2}} \ .
\eea
 The scale of the background energy density is, given the potential \re{V},
\bea
  U^{1/4} &=& \left( \frac{\lambda}{4\xi^2} \right)^{1/4} x \ .
\eea
The Hubble rate during slow-roll inflation is, according to the Friedmann equation,
\bea \label{H}
  H &=&  \left( \frac{\lambda}{12 \xi^2} \right)^{1/2} x^2 \ .
\eea

\begin{figure}[t!]
\centering
\includegraphics[scale=0.28]{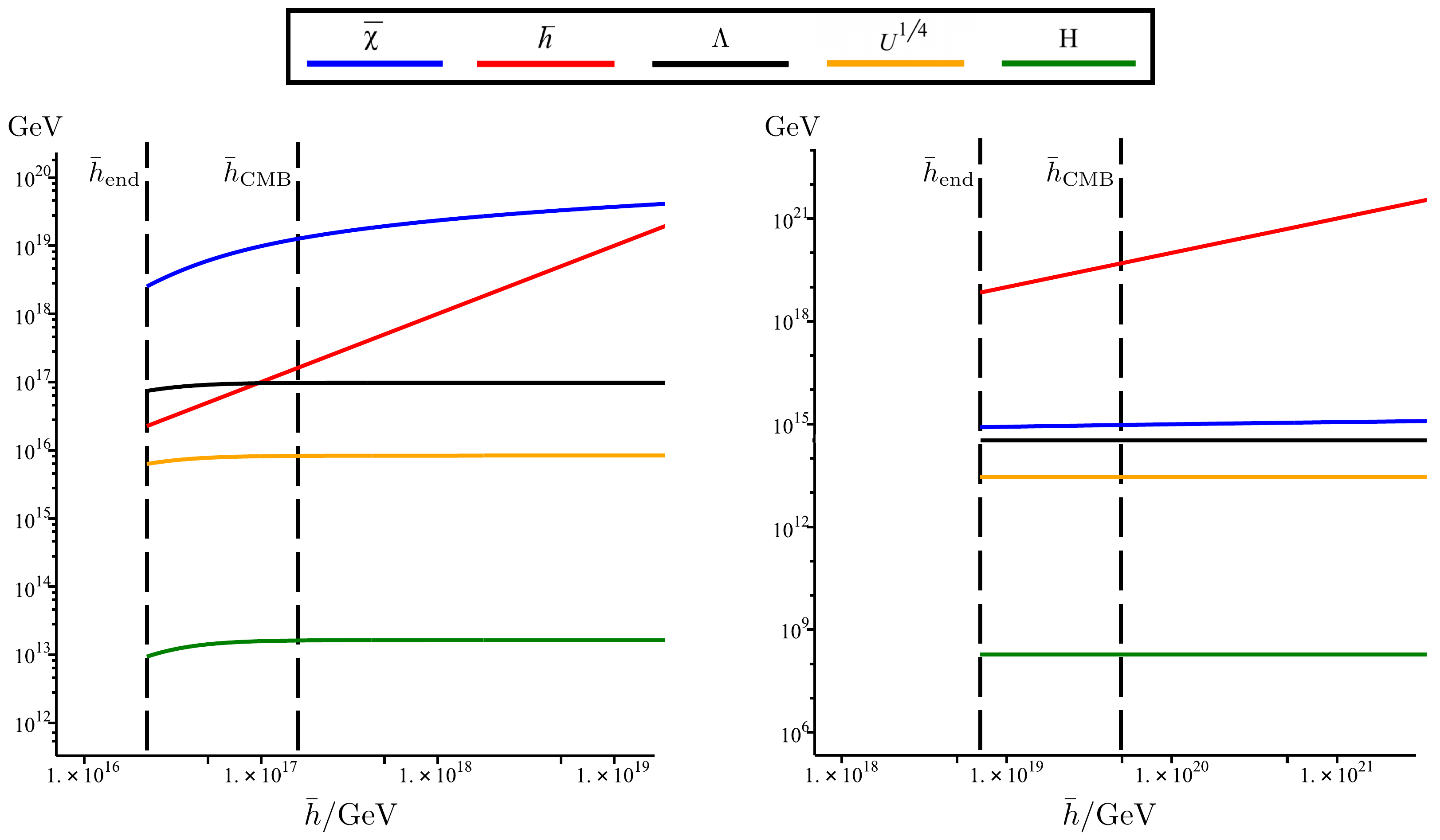}
\caption{
Different scales as a function of the background field $\bh$ in the metric formulation (left) and the Palatini formulation (right). The blue line is the canonical background field $\bar{\chi}$, the red line is the background field $\bh$, the yellow line is the energy density scale $U^{1/4}$, the black line is the unitarity violation scale $\Lambda$, and the green line is the Hubble rate $H$. The dashed vertical lines labelled $h_\textrm{end}$ and $h_\textrm{CMB}$ mark the end of inflation and the CMB pivot scale, respectively.
}
\label{fig:h}
\end{figure}

In \fig{fig:h} we have taken $\lambda=0.13$, the same value as in the EW vacuum, and solved for $\xi$ using \re{As} and \re{N}. We show the unitarity violation scale $\Lambda$, $\bar\chi$, $U^{1/4}$, and $H$ as a function of $\bh$ in the metric and the Palatini formulation. After the end of slow-roll the derivatives of the background field cannot necessarily be neglected, and our approximation ceases to be reliable. The field value $\bh_\text{end}$ at the end of slow-roll is marked on the plot with a vertical line. In the metric formulation it is $\bh_\text{end}=\sqrt{4/(3\xi)}$ \cite{Bezrukov:2007}, and in the Palatini formulation it is $\bh_\text{end}=\sqrt{8}$ \cite{Bauer:2008, Rasanen:2017}. The field value at the CMB pivot scale $k=0.05$ Mpc$^{-1}$ is also marked. In the metric formulation it is $\bh_\text{CMB}=\sqrt{4N/(3\xi)}\approx8/\sqrt{\xi}$, and in the Palatini formulation it is $\bh_\text{CMB}=\sqrt{8N}\approx20$.

In the metric case, the observed CMB amplitude \re{As} together with the value $\lambda=0.13$ gives $\xi=1.6\times10^4$. During inflation, the unitarity violation scale $\Lambda$ is proportional to $1/\sqrt{\xi}$. The Hubble rate $H$ is proportional to $1/\xi$, so it is well below the unitarity violation scale. At the CMB pivot scale, $\Lambda=0.97\times10^{17}$ GeV$=5.1/\sqrt{\xi}$ and $H=1.6\times10^{13}$ GeV. The potential energy scale $U^{1/4}=0.82\times10^{16}$ GeV is an order of magnitude below the cutoff. The original non-minimally coupled Higgs background field $\bh=1.7\times10^{17}$ GeV in the Jordan frame is just above the cutoff, and the minimally coupled canonical field $\bar\chi=1.3\times10^{19}$ GeV is well above it. Note that caution may be required in comparing a Jordan frame field value with an Einstein frame energy scale.

In the Palatini case, $H$ and $U^{1/4}$ have the same dependence on $\xi$ as in the metric case. However, because the amplitude \re{As} is proportional to $1/\xi$ (instead of $1/\xi^2$ as in the metric case), the value $\xi=1.3\times10^9$ is larger. This brings the numerical values of the Hubble rate and the energy density down to $H=4.6\times10^8$ GeV and $U^{1/4}=2.8\times10^{13}$ GeV. The unitarity violation scale is also lower than in the metric case, $\Lambda=3.3\times10^{14}$ GeV$=4.9/\sqrt{\xi}$. As we have calculated the scattering on a Minkowski background, the condition $\Lambda\gg H$ is a requirement for the analysis to be valid, and it is well satisfied in both the metric and the Palatini case. The canonical field value is $\bar\chi=0.94\times10^{15}$ GeV, slightly above $\Lambda$, and the Jordan frame field $\bh=4.8\times10^{19}$ GeV is much higher.

As discussed in \cite{Enckell:2020lvn}, none of the above scales is directly related to scattering processes, which play no role during inflation. The variance of the field is $H/(2\pi)$, which is below (although close to) the cutoff, but it is not clear why this or the other scales should be directly compared to scattering energy. During preheating --where our calculation would have to be extended, as the slow-roll assumption does not hold and gradients can be large-- scattering is relevant, and production of modes with energies above the unitarity violation scale can be a problem \cite{Ema:2016, DeCross:2016, Sfakianakis:2018lzf, Hamada:2020kuy}. Another issue is that non-renormalisable terms in the potential are expected to become important at field values $\sim1/\sqrt{\xi}$, so even if the classical theory is valid, quantum effects can spoil it. The Palatini case is particularly sensitive to higher order terms \cite{Jinno:2019und}. As the same loop factors determine the running of the coupling constants in the scattering amplitudes and in the effective potential, even when scattering processes are not directly relevant for inflation, they can indicate when the inflationary calculation may no longer be trusted. However, the tree-level unitarity violation scale does not necessarily coincide with the true unitarity violation scale nor the scale of new physics \cite{Aydemir:2012nz, Calmet:2013}.

In previous work on unitarity in Higgs inflation in the metric formulation \cite{Barbon:2009, Burgess:2009, Bezrukov:2009db, Burgess:2010zq, Lerner:2009na, Lerner:2010mq, Hertzberg:2010, Bezrukov:2010, Atkins:2010yg, Bezrukov:2011a, Calmet:2013, Weenink:2010, Lerner:2011it, Xianyu:2013, Prokopec:2014, Ren:2014, Escriva:2016cwl, Fumagalli:2017cdo, Mikura:2021clt, Bezrukov:2013, Rubio:2018} it has been argued that the unitarity violation scale in the inflationary large field background is lifted from the EW value $\sim1/\xi$ either to $\sim1/\sqrt{\xi}$ \cite{Atkins:2010yg, Ren:2014, Antoniadis:2021axu, Mikura:2021clt} or to $\sim 1$ \cite{Bezrukov:2010, Bezrukov:2011a, Bezrukov:2013, Rubio:2018} (in the Einstein frame). These arguments are often based on expanding the non-minimal coupling term in the excitations of the fields and using dimensional analysis, but explicit $2\to2$ scattering amplitudes were calculated in \cite{Ren:2014, Antoniadis:2021axu}. Our work agrees with the conclusion that the tree-level unitarity violation scale is lifted to $\sim1/\sqrt{\xi}$.

\begin{figure}[t!]
\centering
\includegraphics[scale=0.28]{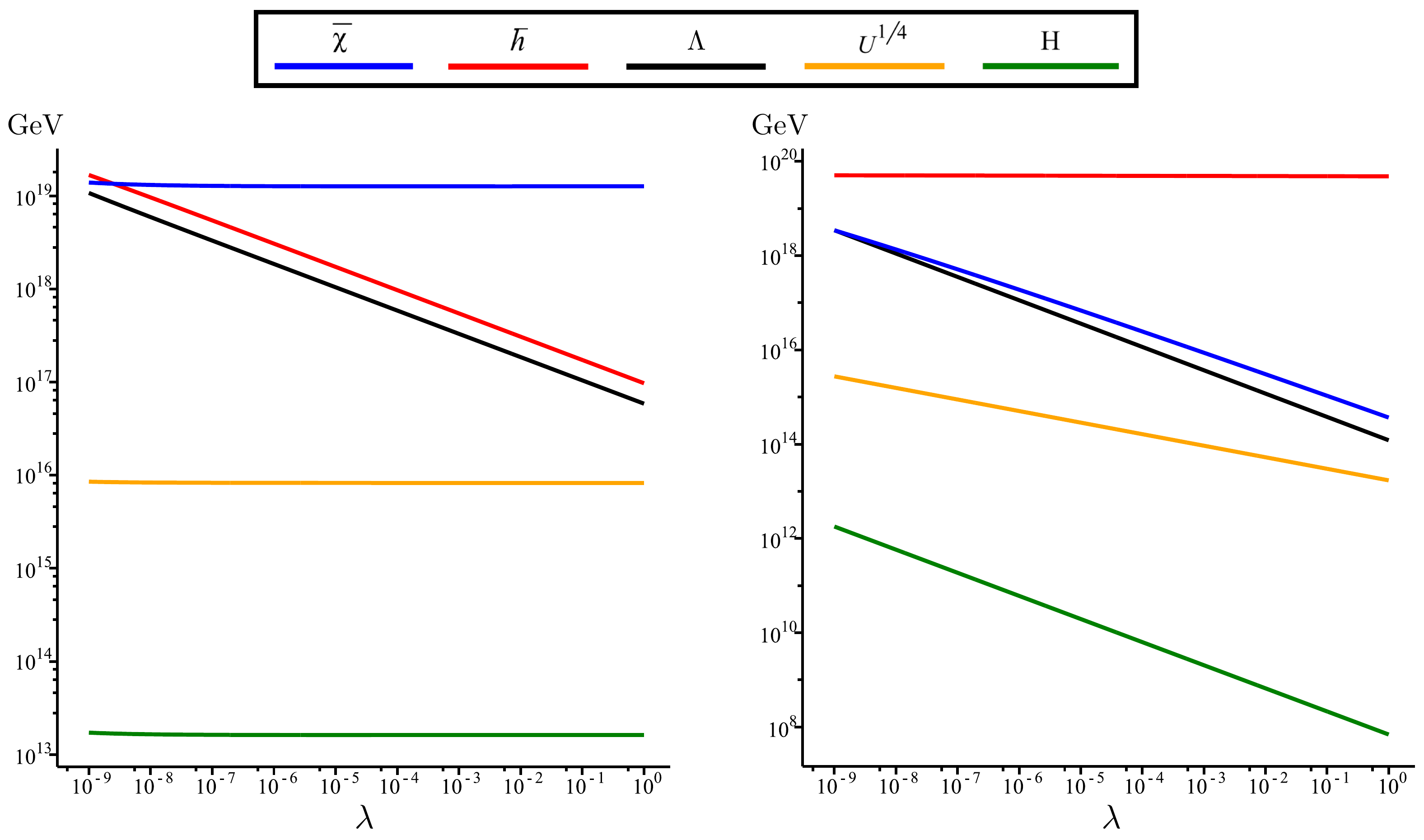}
\caption{
Different scales at the CMB pivot scale as a function of the Higgs quartic coupling $\lambda$ in the metric formulation (left) and the Palatini formulation (right). The line styles are the same as in \fig{fig:h}.
}
\label{fig:lambda}
\end{figure}

In the above discussion, we fixed $\lambda$ to its EW value. However, $\lambda$ runs with scale, and can change significantly between the EW vacuum and the inflationary region \cite{Espinosa:2015a, Espinosa:2015kwx, Iacobellis:2016, Espinosa:2016nld, Hoang:2020iah}. (In contrast, the running of $\xi$ is small.) In \fig{fig:lambda}, we show the effect of having a different value of $\lambda$ at the CMB pivot scale, from 1 down to $10^{-9}$. We fix $\bh$ using \re{N} and {\re{Nh}, and use \re{As} to express $\xi$ in terms of $\lambda$, so that $\Lambda$, $\bh$, $\bar\chi$ and $H$ become functions of $\lambda$. In the Standard Model, $\lambda$ decreases with scale, and can cross zero. It can thus in principle be arbitrarily small at the inflationary scale, but smaller values require more tuning.

In the metric case, matching the observed CMB amplitude $A_s$ \re{As} fixes $\xi=4.3\times10^4\sqrt{\lambda}$, so decreasing $\lambda$ makes $\xi$ smaller, pushing the unitarity violation scale higher. The scales $H$ and $U^{1/4}$ are almost constant, because their amplitude is proportional to the same combination $\lambda/\xi^2$ as $A_s$. They remain below $\Lambda$ for all values of $\lambda$. The value of the Jordan frame field at the CMB scale changes as $\bar h\propto\lambda^{-1/4}$, and the Einstein frame canonical field value there is almost independent of $\lambda$, and both remain above $\Lambda$. Although the plot covers only one value of $\bar h$, the behaviour is typically rather similar across the inflationary region because $\lambda$ changes slowly, unless the coupling is tuned to produced an inflection point or other feature \cite{Allison:2013uaa, Bezrukov:2014bra, Hamada:2014iga, Bezrukov:2014ipa, Rubio:2015zia, Fumagalli:2016lls, Enckell:2016xse, Bezrukov:2017dyv, Masina:2018ejw, Salvio:2017oyf, Ezquiaga:2017fvi, Rasanen:2018a, Enckell:2018a}.

In the Palatini case, the CMB amplitude instead fixes $\xi=1\times10^{10}\lambda$, so now $H\propto\lambda^{-1/2}$, $U^{1/4}\propto\lambda^{-1/4}$. The Jordan frame background field $\bar h$ is almost constant, and the Einstein frame field changes approximately as $\bar\chi\propto\lambda^{-1/2}$. All scales thus grow as $\lambda$ becomes smaller, and $U^{1/4}$ decreases relative to $\Lambda$. The canonical field $\bar\chi$ remains close to and above $\Lambda$. Again, unless $\lambda$ is tuned to produce a feature \cite{Rasanen:2017, Enckell:2018a, Rasanen:2018a, Enckell:2020lvn}, we would expect this behaviour to hold true everywhere in the inflationary region.

\begin{figure}[t!]
\centering
\includegraphics[scale=0.25]{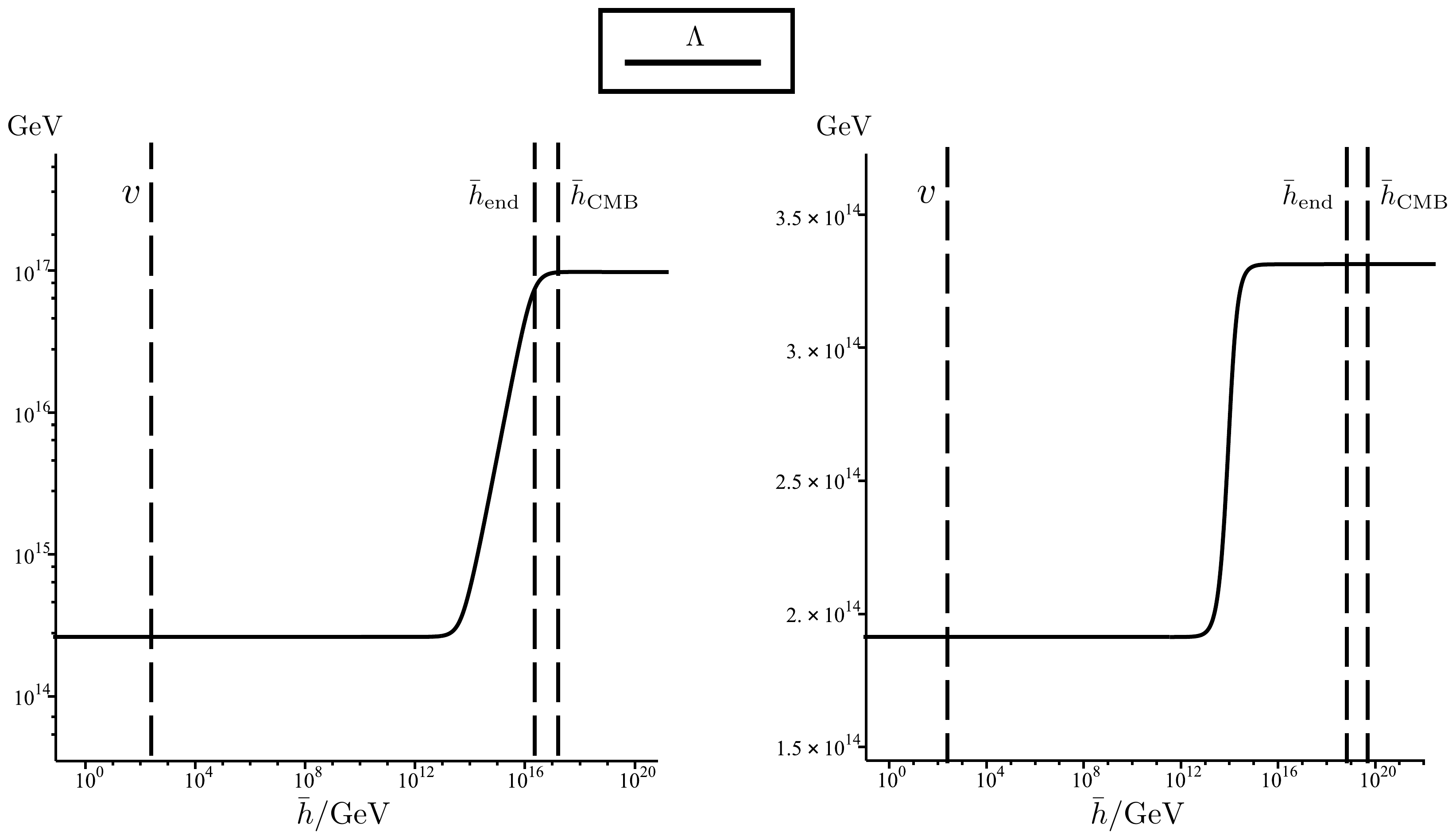}
\caption{
The tree-level unitarity violation scale $\Lambda$ as a function of the Jordan frame background field $\bh$ in the metric formulation (left) and the Palatini formulation (right). The vertical lines mark the Higgs EW vacuum expectation value $v$, the end of inflation, and the CMB pivot scale.
}
\label{fig:cutoff}
\end{figure}

In \fig{fig:cutoff} we show the unitarity violation scale $\Lambda$ as a function of $\bar h$, from the EW scale $v$ up to the CMB pivot scale $h_\text{CMB}$. In the Palatini formulation --in the slow-motion and small fluctuation limit we consider-- $\Lambda$ changes remarkably little, by less than a factor of 2 between the inflationary large field region and the EW minimum. In contrast, in the metric case $\Lambda$ changes by two orders of magnitude between the CMB pivot scale and the EW scale. Between the end of inflation and the end of preheating the derivatives of the field can be large, and our assumption that they can be neglected is not necessarily valid, so the curve may not be accurate there. For small values of $\Lambda$ we would also have to take the masses of the fields into account.

\section{Conclusions} \label{sec:conc}

\para{Unitarity violation at high and low scales.}

We have calculated the tree-level amplitudes for electrically neutral $2\to2$ scattering in the vanilla Higgs inflation model where the Higgs doublet has a non-minimal coupling $\xi$ to the Ricci scalar $R$. We have considered a general slowly varying field background in both the metric formulation and in the Palatini formulation. Using the partial wave unitarity condition, we find the scattering energy $\Lambda$ at which tree-level unitarity is violated. We have also calculated a bound on $\xi$ in the Palatini formulation from the measured Higgs global signal strength, which agrees well with a naive estimate based on the unitarity condition.

We confirm previous estimates  \cite{Atkins:2010yg, Ren:2014, Mikura:2021clt, Bezrukov:2010, Bezrukov:2011a, Bezrukov:2013, Rubio:2018, Antoniadis:2021axu}. In the metric case, $\Lambda\sim\Mpl/\xi\sim10^{14}$ GeV at the EW scale and $\Lambda\sim\Mpl/\sqrt{\xi}\sim10^{17}$ GeV in the inflationary large field background. In the Palatini case, $\Lambda\sim\Mpl/\sqrt{\xi}\sim10^{14}$ GeV changes by less than a factor of two between the EW scale and the inflationary large field background.

Other quantities, such as the Hubble rate $H$, potential energy density scale $U^{1/4}$, Jordan frame background field $\bar h$, and Einstein frame canonical field $\bar\chi$ scale differently with $\xi$ in the metric and in the Palatini formulation. In both the metric and the Palatini formulation, the Hubble rate is orders of magnitude below the tree-level unitarity violation scale (four in the metric case, six in the Palatini case), and $U^{1/4}$ is an order of magnitude below. In the metric case, $\bar\chi$ is two orders of magnitude above $\Lambda$, and $\bar h$ is just above it. In the Palatini case, this is reversed: $\bar\chi$ is slightly above $\Lambda$, and $\bar h$ is five orders of magnitude higher. The differences highlight the interplay between the formulation of general relativity and particle physics scattering, enhanced by the feature that the large non-minimal coupling $\xi$ essentially lowers the gravity scale down from $\Mpl$.

We have assumed that the background field is in slow-roll and smooth, so that its derivatives can be neglected. This means that the results cannot be straightforwardly applied to preheating, where time variation can be rapid and gradients large. Our Palatini case result has the caveat that we neglected fermions, which could potentially lead to unitarity violation at a lower scale. (In the metric case they are subdominant.) Another issue is that higher order scattering amplitudes \cite{Atkins:2010yg, Escriva:2016cwl} could violate unitarity at a lower scales.

It is not clear which (if any) of the above scales should be compared to the unitarity violation scale derived from scattering arguments. In earlier work, it has been argued that the consistency of the theory is in doubt unless the relevant scales (in the simplest argument, the field value $\bh$ during inflation) are smaller than the field value $\Mpl/\xi$ (in the metric case) and $\Mpl/\sqrt{\xi}$ (in the Palatini case) at which new non-renormalisable terms in the potential are expected to be generated by quantum corrections \cite{Barbon:2009, Burgess:2009, Hertzberg:2010}.

Smaller quartic coupling $\lambda$ at the inflationary scale due to running would imply a smaller $\xi$, ameliorating the problem. It is also possible to get a smaller $\xi$ by changing the inflationary dynamics by tuning the running of the Higgs quartic coupling $\lambda$ to create a critical point, hilltop or another feature at the inflationary scale \cite{Allison:2013uaa, Bezrukov:2014bra, Hamada:2014iga, Bezrukov:2014ipa, Rubio:2015zia, Fumagalli:2016lls, Enckell:2016xse, Bezrukov:2017dyv, Masina:2018ejw, Salvio:2017oyf, Ezquiaga:2017fvi, Rasanen:2017, Enckell:2018a, Rasanen:2018a, Enckell:2020lvn}. In the Palatini case, the action can also be changed by including new non-metricity or torsion terms \cite{Rasanen:2018b, Langvik:2020, Shaposhnikov:2020frq}. In the metric case, including a $R^2$ term will lift the unitarity violation scale \cite{Barbon:2015, Salvio:2015kka, Salvio:2017oyf, Kaneda:2015jma, Calmet:2016fsr, Wang:2017fuy, Ema:2017rqn, Pi:2017gih, He:2018gyf, Gorbunov:2018llf, Ghilencea:2018rqg, Wang:2018kly, Gundhi:2018wyz, Karam:2018, Kubo:2018, Enckell:2018c, Ema:2019, Canko:2019mud, Enckell:2018a, He:2020ivk, Bezrukov:2020txg, Ferreira:2021ctx, Hill:2021psc, Antoniadis:2021axu, Mikura:2021clt}, as will a direct coupling between the curvature and the Higgs kinetic term \cite{Germani:2010gm, Germani:2010ux, Kamada:2010qe, Atkins:2010yg, Kamada:2012se, Kamada:2013bia, Germani:2014hqa, Escriva:2016cwl, Fumagalli:2017cdo, Sato:2017qau, Granda:2019wip, Sato:2020ghj, Fumagalli:2020ody, Gialamas:2020vto, Karydas:2021wmx}, or modifications that involve higher derivatives of the Higgs field \cite{Koshelev:2020fok}. The simplest possible solution is that the tree-level unitarity violation scale would not correspond to the real unitarity violation scale due to contributions from higher orders in perturbation theory \cite{Aydemir:2012nz, Calmet:2013}.

\acknowledgments

We thank Ignatios Antoniadis, Anthony Guillen and Kyriakos Tamvakis, as well as Jing Ren, Zhong-Zhi Xianyu and Hong-Jian He, for pointing out a crucial error in the first version of this paper. AI would like to thank Toshifumi Noumi and Jiro Soda for useful conversations. AI was in part supported by JSPS KAKENHI Grant Number JP21J00162 and JP17J00216. WK would like to thank the associateship scheme at the International Center for Theoretical Physics (ICTP) for their support. SR thanks Jose Ramon Espinosa and Tommi Markkanen for helpful correspondence.

\bibliographystyle{JHEP}
\bibliography{uni}

\end{document}